\documentclass[preprint]{aastex}
\shorttitle{SBS~0335--052W -- THE LOWEST-METALLICITY STAR-FORMING 
GALAXY}
\shortauthors{Y. I. Izotov et al.}

\begin{document}
\title{SBS~0335--052W: THE LOWEST-METALLICITY STAR-FORMING GALAXY KNOWN}
\author{Yuri I. Izotov}
\affil{Main Astronomical Observatory, National Academy of Sciences of Ukraine,
03680, Kyiv, Ukraine}
\email{izotov@mao.kiev.ua}

\author{Trinh X. Thuan}
\affil{Astronomy Department, University of Virginia,
    Charlottesville, VA 22903}
\email{txt@virginia.edu}


\author{Natalia G. Guseva}
\affil{Main Astronomical Observatory, National Academy of Sciences of Ukraine,
03680, Kyiv, Ukraine}
\email{guseva@mao.kiev.ua}

\begin{abstract}
We present 4-meter Kitt Peak telescope and 6.5-meter MMT spectrophotometry 
of the extremely low-metallicity galaxy SBS 0335--052W, the 
western companion of the blue compact dwarf galaxy SBS 0335--052E.
These observations have been combined with published 10-meter Keck data 
to derive for the brightest region of SBS 0335--052W an oxygen abundance
12 + log O/H = 7.12 $\pm$ 0.03. This  
makes SBS 0335--052W the lowest metallicity star-forming galaxy known in 
the local universe.
Using a Monte Carlo technique, we fit the spectral energy distribution of 
SBS 0335--052W to derive 
the age of the oldest stars contributing to its optical light. We find that
star formation in SBS 0335--052W began less than 500 Myr ago, 
making it a likely nearby young dwarf galaxy.
\end{abstract}
\keywords{galaxies: abundances --- galaxies: irregular --- 
galaxies: evolution --- galaxies: formation
--- galaxies: ISM --- H {\sc ii} regions --- ISM: abundances}

\section {Introduction}

    Since its discovery as one of the
lowest-metallicity star-forming galaxies known \citep{I90}, 
with oxygen abundance 12 + log O/H $\sim$ 7.30 \citep{M92,I97b,I99,TI05}, 
the blue compact dwarf (BCD) galaxy SBS 0335--052E has often been proposed
as a nearby young dwarf galaxy \citep{I90,I97b,T97}.
It has thus been the subject of numerous multi-wavelength studies. 
On the other hand, its companion galaxy SBS 0335--052W, 
at a projected distance 
of 22 kpc from SBS 0335--052E, has attracted much less attention because of 
its lower brightness. It was 
discovered by \citet{P97} as a dwarf emission-line galaxy associated with  
the brightest of the two intensity peaks of 21 cm emission in the 
large (66 by 22 kpc)  H {\sc i} envelope surrounding  
SBS 0335--052E, with which the other H {\sc i} peak is associated  
 \citep{P01}. 

There has only been one 
determination thus far of the oxygen abundance in SBS 0335--052W, that 
by \citet{L99}. They found that SBS 0335--052W consists of three regions of
star formation and derived 12 + log O/H = 7.22 $\pm$ 0.03 and 7.13 $\pm$ 0.07
for the two brighter regions. Thus, the 
oxygen abundance in SBS 0335--052W was
found to be lower than in SBS 0335--052E and about 
the same as that in the most metal-deficient BCD known,  
I Zw 18 with 12 +log O/H = 7.17 $\pm$ 0.01 and 7.22 $\pm$ 0.01 for its
NW and SE components, respectively \citep{TI05}. 
\citet{L99} discussed the evolutionary status of SBS 0335--052W
and concluded that the galaxy is young, experiencing one of its first
episodes of star formation. 
In this paper we present new observations of SBS 0335--052W 
to check and improve on 
its heavy element abundance determination. The observations and
data reduction are discussed in Sect. \ref{S2}. The element abundances are
derived in Sect. \ref{S3}. The evolutionary status of SBS 0335--052W is 
discussed in Sect. \ref{S4}. Our main conclusions are given in
Sect. \ref{S5}.

\section {Observations and Data Reduction \label{S2}}

We have obtained spectrophotometric observations 
of SBS 0335--052W with the Kitt Peak 4-meter 
telescope\footnote{Kitt Peak National Observatory is operated by the 
Association of Universities for Research in Astronomy, Inc., under contract 
with the National Science Foundation.} and the 
6.5-meter MMT\footnote{MMT Observatory is a joint facility 
of the Smithsonian Institution and the University of Arizona.}.
The Kitt Peak observations were obtained 
on the night of 
2003 January 1.
They were made with the Ritchey-Chr\'etien spectrograph 
in conjunction with a 2048$\times$2048 CCD detector. We use a 
2\arcsec$\times$300\arcsec\ slit with the KPC-10A grating in first
order, and with a GG 375 order separation filter. 
The above instrumental set-up gave a spatial scale
along the slit of 0\farcs69 pixel$^{-1}$, a scale perpendicular to the slit
of 2.7\AA\ pixel$^{-1}$, a spectral range of 3600--7400\AA\ and a spectral
resolution of $\sim$ 7\AA\ (FWHM). 
The total exposure time was 
60 minutes, broken up into three 20 min subexposures. 
The MMT observations
were obtained on 2005 February 4. 
The observations were made with a 
2\arcsec$\times$300\arcsec\ slit and the 800 grooves/mm grating in first
order. This instrumental set-up gave a spatial scale
along the slit of 0\farcs6 pixel$^{-1}$, a scale perpendicular to the slit
of 0.75\AA\ pixel$^{-1}$, a spectral range of 3200--5200\AA\ and a spectral
resolution of $\sim$ 3\AA\ (FWHM). 
The total exposure time was 
45 minutes, broken up into three 15 minutes subexposures. 

\begin{figure}
\rotate
\figurenum{1}
\epsscale{1.0}
\plotone{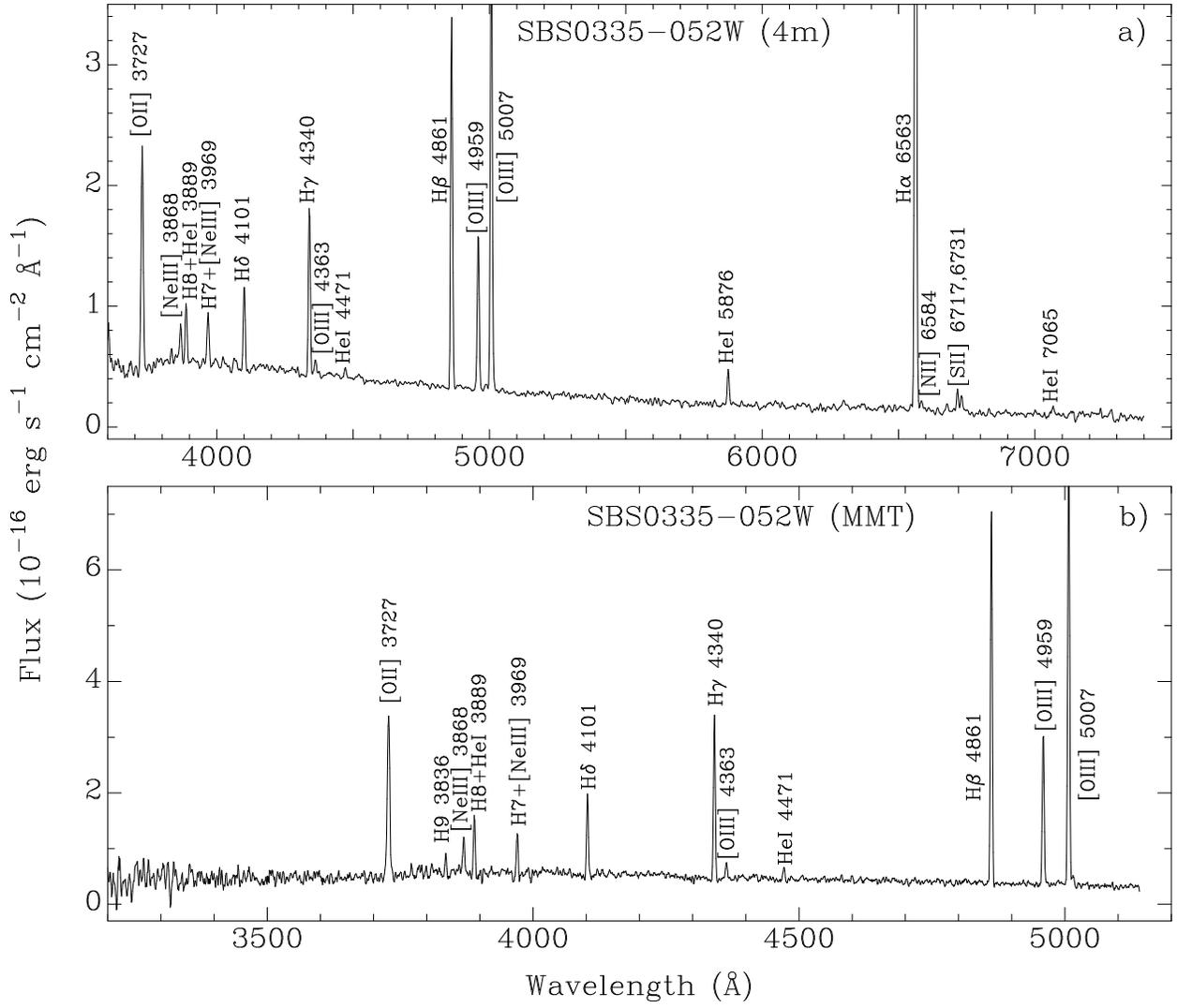}
\caption{Spectrum of SBS 0335--052W
obtained with (a) the 4m Kitt Peak telescope and (b) the MMT.}
\label{Fig1}
\end{figure}

    The two-dimensional spectra were reduced
using IRAF\footnote{IRAF is distributed by National Optical Astronomical 
Observatory, which is operated by the Association of Universities for 
Research in Astronomy, Inc., under cooperative agreement with the National 
Science Foundation.}. 
Since the seeing of $\sim$ 1\arcsec\ -- 2\arcsec\ at both the 
Kitt Peak 4m telescope and the MMT was not good enough
to obtain separate spectra for each of the two brightest star-forming regions 
in SBS0335--052W, we extracted a one-dimensional spectrum of 
only the brightest region within a 6\farcs9$\times$2\arcsec\ 
aperture for the 4m spectrum and within a  
6\arcsec$\times$2\arcsec\ aperture for the MMT spectrum.
The resulting redshift-corrected spectra of SBS 0335--052W are shown in 
the left panels of Figure \ref{Fig1}.
The observed line fluxes were corrected for both reddening \citep{W58} 
and underlying hydrogen stellar absorption.
The corrected fluxes $I$ of the lines used for the extinction and abundance
determination, extinction
coefficients $C$(H$\beta$), equivalent widths EW(H$\beta$) of the H$\beta$
emission line and equivalent widths EW(abs) of the hydrogen
absorption stellar lines are given in Table \ref{Tab1}.
The line flux errors listed include 
statistical errors derived from non-flux calibrated spectra, in 
addition to errors introduced in the standard star absolute flux calibration, 
which we set to 1\% of the line fluxes. These errors will be later propagated 
into the calculation of abundance errors.

\section {Physical Conditions and Element Abundances \label{S3}}

   To determine element abundances, we follow generally 
the procedures of \citet{ITL94,ITL97} and \citet{TIL95}.
We adopt a two-zone photoionized H {\sc ii}
region model: a high-ionization zone with temperature $T_e$(O {\sc iii}), 
where [O {\sc iii}] and [Ne {\sc iii}] lines originate, and a 
low-ionization zone with temperature $T_e$(O {\sc ii}), where [O {\sc ii}], 
[N {\sc ii}], [S {\sc ii}] and [Fe {\sc iii}] lines originate. 
As for the [S {\sc iii}] lines they originate in the 
intermediate zone between the high and low-ionization regions \citep{G92}. 
The temperature $T_e$(O {\sc iii}) is calculated using the 
[O {\sc iii}] $\lambda$4363/($\lambda$4959+$\lambda$5007) ratio. 
For other ionic temperatures, we use
the relation between the electron temperatures $T_e$(O {\sc iii}) and
$T_e$(O {\sc ii}), and  that between $T_e$(O {\sc iii}) and 
$T_e$(S {\sc iii}) obtained by \citet{I05} from
the H {\sc ii} photoionization models of \citet{SI03}. These are based on
more recent stellar atmosphere models and improved
atomic data as compared to \citet{S90} models. 
The [S {\sc ii}] 
$\lambda$6717/$\lambda$6731 ratio is used to determine the electron density 
$N_e$(S {\sc ii}) for the 4-meter data. As [S {\sc ii}] emission 
lines were not in the observed wavelength region of the MMT spectrum, 
$N_e$(S {\sc ii}) was set to 10 cm$^{-3}$
for those data. Ionic and total heavy element abundances are derived 
using expressions for ionic abundances and ionization correction 
factors obtained by \citet{I05}.
The element abundances for the 4-meter and MMT observations
are given in the first two columns of 
Table \ref{Tab2} along with the adopted electron temperatures for
different ions and electron number densities.

The derived oxygen abundance in the
brightest region of SBS 0335--052W is extremely low, 12 + log O/H = 
7.13 $\pm$ 0.08 and 7.11 $\pm$ 0.05 for the 4-meter telescope and MMT
observations respectively, with a weighted mean of 7.12 $\pm$ 0.04.
This value is 
lower than the oxygen abundance 
of 7.31 $\pm$ 0.01 in SBS 0335--052E, and of 
7.17 $\pm$ 0.01 and 7.22 $\pm$ 0.01 respectively in the NW and SE components 
of I Zw 18 recently derived by \citet{TI05}.
Our new value of the oxygen abundance
in the brightest region of SBS 0335--052W is 
significantly different from the
one of 7.22 $\pm$ 0.03 derived by \citet{L99} for the same region. It
is similar, however, to the oxygen abundance of 7.13 $\pm$ 0.08
derived by \citet{L99} for the fainter region. 

To resolve the discrepancy, we have reanalyzed the 10-meter
Keck\footnote{The data presented herein were obtained at the W.M. Keck 
Observatory, which is operated as a scientific partnership among the 
California Institute of Technology, the University of California and the 
National Aeronautics and Space Administration. The Observatory was made 
possible by the generous financial support of the W.M. Keck Foundation.} 
spectrum of \citet{L99}.
We have remeasured the emission-line
fluxes in both the bright and faint regions (Table \ref{Tab1}) 
and derived respectively 12 + log O/H = 7.23 $\pm$ 0.04 and 7.12 $\pm$ 0.11,
in very good agreement with those derived by \citet{L99}.
The discrepancy is thus not caused by faulty line flux measurements.
We next inspected the profiles of different emission lines in the
Keck spectrum of the bright region. We found that while the profiles of other
lines are symmetric and broad, the profile of [O {\sc iii}] $\lambda$4363 
is asymmetric and narrower as if its red wing has been clipped off. 
The FWHM of this
emission line is $\sim$ 30\% narrower than that of the nearest and comparable
in flux He {\sc i} $\lambda$4471 emission line. 
The discrepancy is  thus 
apparently due to problems with the [O {\sc iii}] $\lambda$4363 emission line 
flux in the Keck spectrum. If the flux of this line
is increased by 30\%, then the oxygen abundance
in the bright region is 7.12 $\pm$ 0.04, in excellent agreement with the 
oxygen abundances derived from other observations (Table \ref{Tab2}).

The weighted mean oxygen abundance for the bright region 
in SBS 0335--052W is thus 7.12 $\pm$ 0.04
if only the 4-meter telescope and MMT observations 
are included, and 7.12 $\pm$ 0.03 if the re-examined Keck observations of 
both bright and faint regions
are added. This makes SBS 0335--052W the lowest-metallicity
star-forming galaxy known in the local universe. Its metallicity is 
slightly lower than that of I Zw 18, 
by 0.05 dex (at the $\sim$ 2$\sigma$ level) for its NW component,
and by 0.10 dex (at the $\sim$ 3$\sigma$ level) for its SE component. 
The other heavy elements to oxygen abundance ratios in SBS 0335--052W 
(Table \ref{Tab2}) are in very good agreement with the mean ratios for
the most metal-deficient BCDs \citep{TIL95,IT99,I05}. 

It is important to 
note that the differences in metallicities between the three 
most metal-deficient gas-rich star-forming galaxies known in the local 
universe, SBS 0335--052 W, I Zw 18 and SBS 0335--052E, 
are very small, at the level of a few hundredths of a dex. 
Thus their ionized gaseous component 
 is enriched at practically the same level, and on the basis of   
these three BCDs, there appears to exist
 a lowest observable metallicity of about 2\% that of the Sun
 in star-forming galaxies, set by the pollution of the ionized gas by the 
current starburst (Kunth \& Sargent 1986).

\section{The Evolutionary Status of SBS 0335--052W \label{S4}}

     The extremely low metallicity of SBS 0335--052W suggests that it 
may be a young galaxy. However, this galaxy 
at a distance of 54.3 Mpc 
\citep[e.g.,][]{TI97} is too far to be resolved into individual stars by 
{\sl HST}. 
Therefore, the only way to derive its age is by studying its integrated 
spectroscopic and photometric properties. 

We have carried out a series of 
Monte Carlo simulations of the stellar populations in 
the bright region of SBS 0335--052W
to reproduce the 4-meter, MMT and Keck
spectral energy distributions (SEDs). 
We have used the Padua stellar evolution
models \citep{Gi00}\footnote{http://pleiadi.pd.astro.it.} with a heavy
element mass fraction $Z$ = 0.0004, corresponding to the gaseous 
oxygen abundance in SBS 0335--052W. 
We have adopted a stellar initial mass function with a Salpeter slope, an
upper mass limit of 100 $M_\odot$ and a lower mass limit of 0.1 $M_\odot$.
For the star formation history of SBS 0335--052W, we have  
considered two short bursts of star formation: a recent 
burst with age $t$(young) varying between 0.5 and 8 Myr which accounts for the 
young stellar population and a prior burst 
with age $t$(old) varying 
between 20 Myr and 15 Gyr and responsible for the older stars. 
CMD studies of star-forming dwarf 
galaxies  \citep[e.g., ][]{IT04} suggest that 
such a simple 2-burst model can reproduce adequately the 
main features of the stellar populations in these dwarfs. 
The contribution of each burst
to the SED is defined by the ratio of the masses of stellar populations
formed respectively 
in the old and young bursts, $b$ = $M$(old)/$M$(young), which we vary 
between 0.01 and 100. The contribution of 
gaseous emission is determined from the
observed equivalent width (EW) of the H$\beta$ emission line. The 
fraction of gaseous continuum to total light depends on the adopted 
electron temperature $T_e$(H$^+$) in the H$^+$ zone, since EW(H$\beta$)
for pure gaseous emission decreases with increasing $T_e$(H$^+$).
Given that $T_e$(H$^+$) is not necessarily
equal to $T_e$(O {\sc iii}), we chose to vary it in the range 
(0.8 -- 1.2)$\times$$T_e$(O {\sc iii}). We assume that
the extinction for the stellar light $C$(stars) is close to 
that for the ionized gas, $C$(gas).
Both $C$(stars) and $C$(gas) are varied in the narrow range 
(0.95 -- 1.05)$\times$$C$(H$\beta$), where 
$C$(H$\beta$) is the extinction derived 
from the hydrogen Balmer decrement (Table \ref{Tab1}).
We then run 10$^6$ Monte Carlo models 
varying simultaneously $C$(gas), $C$(stars), 
$t$(young), $t$(old), $b$ and $T_e$(H$^+$).  
To judge the goodness of each model's fit to the observed SED, 
we have computed a $\chi^2$ for each Monte Carlo realization. 

\begin{figure}
\rotate
\figurenum{2}
\epsscale{0.48}
\hspace*{0.0cm}\plotone{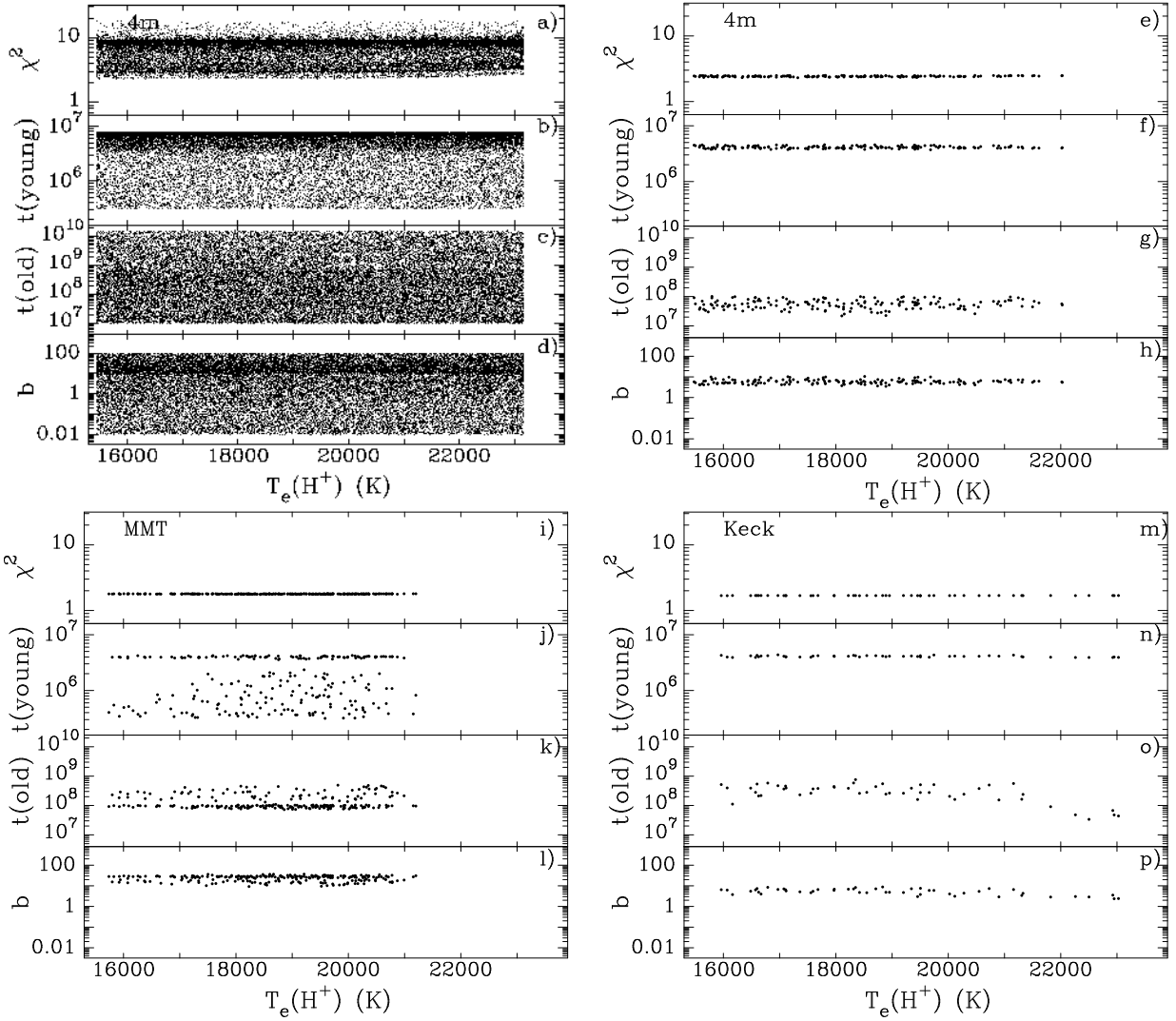}
\hspace*{0.2cm}\plotone{f2b.eps}
\hspace*{0.0cm}\plotone{f2c.eps}
\hspace*{0.2cm}\plotone{f2d.eps}
\caption{(a) -- (d) Parameter space explored in Monte Carlo
simulations to fit the 4m spectrum. The parameters that are varied are: 
 $t$(old) and $t$(young), respectively the ages
of the old and young stellar populations, $b$ the ratio of their masses,
and $T_e$(H$^+$), the electron temperature in the H$^+$ zone. 
$\chi^2$ is an estimator of the goodness of fit.
(e) -- (h) Parameter distribution for the best fits to the 4m spectrum, 
characterized by the
lowest $\chi^2$s in (a) -- (d). (i) -- (l) Same as (e) -- (h) but for 
the MMT spectrum. (m) -- (p) Same as (e) -- (h) but for the Keck spectrum. }
\label{Fig2}
\end{figure}

\begin{figure}
\rotate
\figurenum{3}
\epsscale{1.0}
\plotone{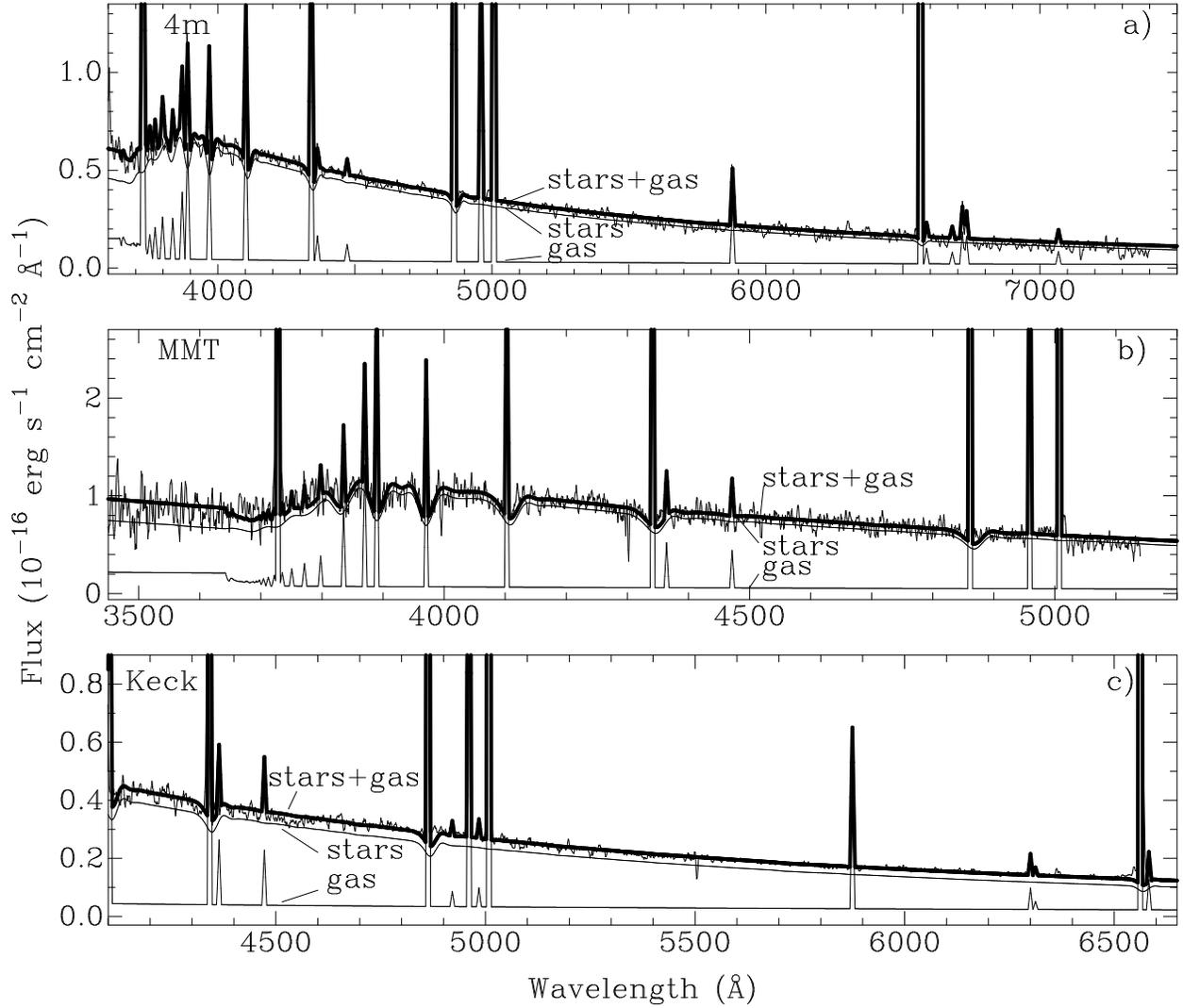}
\caption{Best fits of model SEDs 
to the redshift- and extinction-corrected 
(a) 4m telescope, (b) MMT and (c) Keck spectra are shown by thick solid lines. 
The separate contributions 
from the stellar and ionized gas components are shown
by thin solid lines.}
\label{Fig3}
\end{figure}

In Fig. \ref{Fig2}a -- \ref{Fig2}d, we show the parameter space explored
in 37,581 Monte Carlo realizations to fit the 4-meter spectrum, along 
with the $\chi^2$s of these realizations. We do not show $C$(gas) and 
$C$(stars) as they vary in the narrow range of $\pm$5\% around $C$(H$\beta$).
The remaining realizations out of the 10$^6$ models run are not displayed 
in Fig. \ref{Fig2}a -- \ref{Fig2}d because they predict a EW(H$\beta$) 
which deviates by more than 10\% from the observed one.
 It is seen that the age
$t$(old) of the old stellar population, the age $t$(young) of the young
stellar population and the ratio $b$ of their masses vary in a wide range.
However, if only the best solutions with the lowest $\chi^2$s are considered
(Fig. \ref{Fig2}e -- \ref{Fig2}h), then the range 
of these parameters is considerably narrowed. 
The best solution is seen to consist of an old stellar population with age 
$\la$ 100 Myr, 
contributing $\sim$ 30\% and $\sim$ 50\% of the total 
stellar light at $\sim$ 4000\AA\ and $\sim$ 7500\AA\ respectively. Its mass 
is $\sim$ 6 times larger than the mass of the 
young stellar population, the age of which is $\sim$ 4 Myr. 
The best fit SED to the 4-meter
spectrum is shown in Fig. \ref{Fig3}a, with the contributions of 
the stellar and gaseous components shown separately.
In the same manner, we have run 10$^6$ models to fit both the MMT and Keck 
spectra. Fig. \ref{Fig2}i -- \ref{Fig2}l and 
Fig. \ref{Fig2}m -- \ref{Fig2}p show the best solutions for the MMT and Keck
spectra respectively, out of 34,408 and 61,177 Monte Carlo realizations
which fit the observed EW(H$\beta$). 
The solutions with the lowest $\chi^2$s give $t$(young) $\la$ 4 Myr 
(MMT and Keck), 
$t$(old) $\la$ 500 Myr (MMT and Keck), $b$ $\sim$ 10 -- 30 (MMT) and 
$b$ $\sim$ 6 (Keck), consistent with the values derived from the 
4m spectrum. 
In the MMT spectrum, the old stellar population contributes 
$\sim$ 30\% of the total stellar light at $\sim$ 4200\AA, while in the Keck 
spectrum this contribution is lower, $\sim$ 10\%. The difference is presumably 
due to the smaller 1\arcsec$\times$180\arcsec\ slit 
used in the Keck observations. Varying  
the stellar IMF slope in the range $\alpha$ = --2 - --3 and the 
lower stellar mass limit 
in the range 0.1 - 1 $M_\odot$ do not change significantly our age estimates.
The best fits of the observed 
MMT and Keck spectra are shown in Figs. \ref{Fig3}b and \ref{Fig3}c.

Our age estimates of the oldest stellar population in SBS 0335--052W 
obtained from fitting the SEDs are in good agreement with those obtained 
by \citet{P04} from
broad-band photometry of the extended low-surface-brightness underlying 
component. Those authors found
that the $U-B$, $B-V$ and $V-I$ colors of the outer regions of SBS 0335--052W
are consistent with an age $\la$ 1 Gyr. Only the $V-R$ color gives a 
larger age. However, age determination from the latter color 
is less reliable as there is a 
significant contamination of the $R$ band by H$\alpha$ gaseous emission.
Thus the good  agreement between our derived ages and those 
derived from photometric data of the outer regions of SBS 0335--052W 
shows that 
the SED fitting technique is quite sensitive to the 
the presence of old stellar
populations, despite the fact that 4m, MMT and Keck spectra cover 
only the small brightest part of SBS 0335--052W.

In summary, 
SBS 0335--052W appears to be a young galaxy with the age of its oldest stars
not exceeding $\sim$ 500 Myr, as in the case of I Zw 18 \citep{IT04}.

\section{Conclusions \label{S5}}

Our new spectroscopic observations of the dwarf star-forming galaxy 
SBS 0335--052W with the 4-meter telescope and the 6.5-meter MMT 
have led us to the following conclusions:

1. The weighted mean oxygen abundance in SBS 0335--052W is 12 + log O/H =
7.12 $\pm$ 0.04 based on present observations, and 
7.12 $\pm$ 0.03 if previous 10-meter Keck observations by \citet{L99} are 
reanalyzed and added. This makes SBS 0335--052W the lowest-metallicity
star-forming galaxy known.

2. Monte Carlo simulations of the spectral energy distributions of 
SBS 0335--052W by varying the ages of the young and old stellar populations
and their mass ratio show that the age of the oldest stars is not greater
than 500 Myr. Thus, SBS 0335--052W is likely a nearby young galaxy.

\acknowledgements

The MMT time was available thanks to a grant from the 
Frank Levinson Fund of the Peninsula Community Foundation 
to the Astronomy Department of the University of Virginia.
The research described in this publication was made possible in part by Award
No. UP1-2551-KV-03 of the U.S. Civilian Research \& Development Foundation 
for the Independent States of the Former Soviet Union (CRDF).
It has also been supported by NSF grant AST-02-05785.
Y.I.I. and N.G.G. thank the hospitality of the Astronomy Department of 
the University of Virginia. 



\clearpage

\begin{deluxetable}{lrrrr}
\tablecolumns{5}
\tablewidth{0pt}
\tablecaption{Emission Line Fluxes \label{Tab1}}
\tablehead{
\colhead{Ion}&\multicolumn{1}{c}{4m}&
\multicolumn{1}{c}{MMT}& 
\multicolumn{2}{c}{Keck} \\ \cline{4-5}
\colhead{}&\colhead{}&\colhead{}&\colhead{bright region}&\colhead{faint region} }
\startdata
 3727\ [O {\sc ii}]        & 70.6$\pm$2.9& 69.4$\pm$1.7&\nodata~~~      &\nodata~~~       \\
 3836\ H9                  &  6.5$\pm$2.1&  7.5$\pm$1.1&\nodata~~~      &\nodata~~~       \\
 3868\ [Ne {\sc iii}]      & 10.7$\pm$1.2& 10.9$\pm$0.7&\nodata~~~      &\nodata~~~       \\
 3889\ He {\sc i}+H8       & 18.5$\pm$1.7& 20.7$\pm$1.1&\nodata~~~      &\nodata~~~       \\
 3969\ [Ne {\sc iii}]+H7   & 19.1$\pm$1.7& 18.2$\pm$0.9&\nodata~~~      &\nodata~~~       \\
 4101\ H$\delta$           & 25.8$\pm$1.5& 26.4$\pm$1.1& 26.6$\pm$1.3& 24.7$\pm$3.9 \\
 4340\ H$\gamma$           & 48.1$\pm$1.5& 47.5$\pm$1.2& 46.2$\pm$1.1& 44.3$\pm$2.8 \\
 4363\ [O {\sc iii}]       &  4.0$\pm$0.8&  4.1$\pm$0.5&  4.4$\pm$0.4\tablenotemark{a}&  4.4$\pm$1.1 \\
 4471\ He {\sc i}          &  2.6$\pm$0.8&  3.4$\pm$0.4&  3.7$\pm$0.3&  3.8$\pm$0.9 \\
 4861\ H$\beta$            &100.0$\pm$2.0&100.0$\pm$1.9&100.0$\pm$1.7&100.0$\pm$2.4 \\
 4959\ [O {\sc iii}]       & 43.0$\pm$1.1& 41.8$\pm$0.9& 44.3$\pm$0.7& 43.8$\pm$1.0 \\
 4988\ [Fe {\sc iii}]      &\nodata~~~      &\nodata~~~      &  1.3$\pm$0.2&\nodata~~~       \\
 5007\ [O {\sc iii}]       &124.1$\pm$2.4&123.2$\pm$2.2&131.3$\pm$2.1&130.2$\pm$2.4 \\
 5876\ He {\sc i}          &  9.1$\pm$0.6&\nodata~~~      &  9.4$\pm$0.2&  9.6$\pm$0.4 \\
 6300\ [O {\sc i}]         &\nodata~~~      &\nodata~~~      &  1.4$\pm$0.1&\nodata~~~       \\
 6312\ [S {\sc iii}]       &\nodata~~~      &\nodata~~~      &  0.5$\pm$0.1&\nodata~~~       \\
 6563\ H$\alpha$           &274.3$\pm$5.3&\nodata~~~      &274.3$\pm$4.6&273.4$\pm$5.2 \\
 6583\ [N {\sc ii}]        &  2.5$\pm$0.7&\nodata~~~      &  2.0$\pm$0.1&  2.1$\pm$0.3 \\
 6678\ He {\sc i}          &  1.9$\pm$0.6&\nodata~~~      &\nodata~~~      &\nodata~~~       \\
 6717\ [S {\sc ii}]        &  5.1$\pm$0.6&\nodata~~~      &\nodata~~~      &\nodata~~~       \\
 6731\ [S {\sc ii}]        &  4.4$\pm$0.7&\nodata~~~      &\nodata~~~      &\nodata~~~       \\ 
 7065\ He {\sc i}          &  2.0$\pm$0.7&\nodata~~~      &\nodata~~~      &\nodata~~~       \\ \\
 $C$(H$\beta$) dex   &\multicolumn {1}{c}{0.06$\pm$0.02}&\multicolumn {1}{c}{0.12$\pm$0.02}&\multicolumn {1}{c}{0.13$\pm$0.02}&\multicolumn {1}{c}{0.00$\pm$0.02} \\
EW(H$\beta$)\ \AA      &\multicolumn {1}{c}{82$\pm$1}&\multicolumn{1}{c}{80$\pm$1}&\multicolumn{1}{c}{105$\pm$1} &\multicolumn{1}{c}{120$\pm$1}\\
EW(abs)\ \AA      &\multicolumn {1}{c}{1.4$\pm$0.4}&\multicolumn{1}{c}{ 0.8$\pm$0.5}&\multicolumn{1}{c}{ 5.9$\pm$0.4} &\multicolumn{1}{c}{ 3.8$\pm$1.8}\\
\enddata
\tablenotetext{a}{The measured flux is 3.4 $\pm$ 0.4. It has been 
increased by 30\% 
as discussed in the text.}
\end{deluxetable}

\clearpage

\begin{deluxetable}{lccccc}
\tablecolumns{6}
\tablewidth{0pt}
\tablecaption{Ionic and Element Abundances \label{Tab2}}
\tablehead{
\colhead{Property}&\colhead{4m}&\colhead{MMT}&\multicolumn{2}{c}{Keck} &\colhead{weighted} \\ \cline{4-5}
\colhead{}&\colhead{}&\colhead{}&\colhead{bright region\tablenotemark{a}}&\colhead{faint region}& \colhead{mean} }
\startdata
$T_e$(O {\sc iii})(K)                                &19400$\pm$2240                   &19770$\pm$1330      &19840$\pm$1030    &19900$\pm$3040 &      \\
$T_e$(O {\sc ii})(K)                                 &15550$\pm$1670                   &15590$\pm$~\,970    &15600$\pm$~\,750  &15600$\pm$2200 &    \\
$T_e$(S {\sc iii})(K)                                &18350$\pm$1860                   &18600$\pm$1100      &18650$\pm$~\,850  &18690$\pm$2520 &    \\
$N_e$(S {\sc ii})(cm$^{-3}$)                         &  330$\pm$100                    &    10         &    10       &    10   &    \\ \\
O$^+$/H$^+$($\times$10$^5$)                          &0.60~~$\pm$~0.16                 &0.56~~$\pm$~0.09    &0.56~~$\pm$~0.07\tablenotemark{b}&0.56~~$\pm$~0.20\tablenotemark{b} \\
O$^{++}$/H$^+$($\times$10$^5$)                       &0.77~~$\pm$~0.20                 &0.72~~$\pm$~0.11    &0.77~~$\pm$~0.09  &0.75~~$\pm$~0.26     \\
O/H($\times$10$^5$)                                  &1.36~~$\pm$~0.26                 &1.28~~$\pm$~0.14    &1.33~~$\pm$~0.11  &1.32~~$\pm$~0.33     \\
12 + log(O/H)                                        &7.13~~$\pm$~0.08                 &7.11~~$\pm$~0.05    &\,~7.12~~$\pm$~0.04\tablenotemark{b}  &\,~7.12~~$\pm$~0.11\tablenotemark{b}  &  7.12~~$\pm$~0.03  \\ \\
N$^+$/H$^+$($\times$10$^7$)                          &1.70~~$\pm$~0.46                 &\nodata             &1.39~~$\pm$~0.12  &1.42~~$\pm$~0.37       \\
ICF(N)                                               &2.30\,~~~~~~~~~~                 &\nodata             &2.38\,~~~~~~~~~~  &2.36\,~~~~~~~~~~     \\ 
log(N/O)                                             &--1.54~~$\pm$~0.14~~             &\nodata             &--1.61~~$\pm$~0.05~~ &--1.59~~$\pm$~0.16~~ &--1.60~~$\pm$~0.04~~ \\ \\
Ne$^{++}$/H$^+$($\times$10$^5$)                      &0.15~~$\pm$~0.04                 &0.14~~$\pm$~0.02    &\nodata    &\nodata              \\
ICF(Ne)                                              &1.19\,~~~~~~~~~~                 &1.19\,~~~~~~~~~~    &\nodata    &\nodata              \\ 
log(Ne/O)                                            &--0.89~~$\pm$~0.20~~             &--0.88~~$\pm$~0.11~~&\nodata    &\nodata  &--0.88~~$\pm$~0.10~~  \\ \\
S$^+$/H$^+$($\times$10$^7$)                          &0.89~~$\pm$~0.16                 &\nodata             &\nodata    &\nodata              \\
S$^{++}$/H$^+$($\times$10$^7$)                       &1.58~~$\pm$~0.40\tablenotemark{c}&\nodata             &\nodata    &\nodata              \\
ICF(S)                                               &0.93\,~~~~~~~~~~                 &\nodata             &\nodata    &\nodata              \\ 
log(S/O)                                             &\,~--1.77~~$\pm$~0.11\tablenotemark{c}~~             &\nodata             &\nodata    &\nodata &--1.77~~$\pm$~0.11~~             \\ \\
Fe$^{+}$/H$^+$($\times$10$^6$)                       &\nodata                          &\nodata             &0.22~~$\pm$~0.03   &\nodata      \\
ICF(Fe)                                              &\nodata                          &\nodata             &3.13\,~~~~~~~~~~   &\nodata      \\ 
log(Fe/O)                                            &\nodata                          &\nodata             &--1.28~~$\pm$~0.08~~ &\nodata& --1.28~~$\pm$~0.08~~
\enddata
\tablenotetext{a}{Electron temperatures and element abundances are derived 
with the corrected flux of the
[O {\sc iii}] $\lambda$4363 emission line. Without that correction, 
the oxygen abundance would have been 12 + log O/H = 7.23 $\pm$ 0.04.}
\tablenotetext{b}{The abundance is calculated from the relative flux 
of the [O {\sc ii}] $\lambda$3727 line derived from the MMT spectrum.}
\tablenotetext{c}{The abundance is calculated from the relative flux 
of the [S {\sc iii}] $\lambda$6312 line derived from the Keck spectrum.}
\end{deluxetable}

\clearpage

\clearpage

\clearpage

\end{document}